\documentclass[fleqn,usenatbib]{mnras}

\usepackage[T1]{fontenc}

\DeclareRobustCommand{\VAN}[3]{#2}
\let\VANthebibliography\thebibliography
\def\thebibliography{\DeclareRobustCommand{\VAN}[3]{##3}\VANthebibliography}

\usepackage{graphicx}
\usepackage{amsmath}	
\usepackage{amssymb}	
\usepackage{bm}
\usepackage{hyperref}
\usepackage[normalem]{ulem}
\usepackage{float}

\usepackage{newtxtext}
\usepackage[varvw]{newtxmath}

\newcommand{\diff}{\ensuremath{\mathrm{d}}}

\newcommand{\lcut}{\lambda_\mathrm{fs}}
\newcommand{\vc}{v_\mathrm{circ}}

\newcommand{\rhoM}{\rho_\mathrm{m}}
\newcommand{\rhoMo}{\rho_{\mathrm{m},0}}
\newcommand{\rhoL}{\rho_\Lambda}
\newcommand{\OmegaM}{\Omega_\mathrm{m}}
\newcommand{\OmegaL}{\Omega_\Lambda}

\newcommand{\zc}{z_\mathrm{c}}

\newcommand{\rh}{r_\mathrm{h}}
\newcommand{\Rh}{R_\mathrm{h}}

\def\chg#1{#1}

\title[Prompt cusps and warm dark matter]{Massive prompt cusps: A new signature of warm dark matter}

\author[M. S. Delos]{
M. Sten Delos\thanks{E-mail: sten@mpa-garching.mpg.de}
\\
Max Planck Institute for Astrophysics, Karl-Schwarzschild-Str. 1, 85748 Garching, Germany
}

\date{Accepted XXX. Received YYY; in original form ZZZ}

\pubyear{2023}

\begin{document}

\label{firstpage}
\pagerange{\pageref{firstpage}--\pageref{lastpage}}
\maketitle

\begin{abstract}
Every dark matter halo and subhalo is expected to have a prompt $\rho\propto r^{-1.5}$ central density cusp, which is a relic of its condensation out of the smooth mass distribution of the early universe. The sizes of these prompt cusps are linked to the scales of the peaks in the initial density field from which they formed. In warm dark matter (WDM) models, the smoothing scale set by free streaming of the dark matter can result in prompt cusps with masses of order $10^7$~M$_\odot$. We show that WDM models with particle masses ranging from 2 to 6~keV predict prompt cusps that could detectably alter the observed kinematics of Local Group dwarf galaxies. Thus, prompt cusps present a viable new probe of WDM. A prompt cusp's properties are highly sensitive to when it formed, so prospects can be improved with a better understanding of when the haloes of the Local Group dwarfs originally formed. Tidal stripping can also affect prompt cusps, so constraints on satellite galaxy orbits can further tighten WDM inferences.
\end{abstract}




\section{Introduction}

The onset of the formation of dark matter haloes in the early universe was marked by the monolithic collapse of smooth peaks in the density field. This process creates \chg{$\rho \propto r^{-1.5}$ density cusps \citep{2010ApJ...723L.195I,2013JCAP...04..009A,2014ApJ...788...27I,2015MNRAS.450.2172P,2017MNRAS.471.4687A,2018MNRAS.473.4339O,2018PhRvD..97d1303D,2018PhRvD..98f3527D,2019PhRvD.100b3523D,2020MNRAS.492.3662I,2021A&A...647A..66C,2022MNRAS.517L..46W}, which form promptly at collapse and} persist largely unaltered through the subsequent growth phases of their haloes \citep{2023MNRAS.518.3509D}.
Consequently, \chg{one such \textit{prompt cusp}} is expected to reside at the centre of every halo and subhalo. The properties of each prompt cusp are closely connected to those of its precursor density peak \citep{2019PhRvD.100b3523D,2023MNRAS.518.3509D}, so prompt cusps carry precise information about the primordial mass distribution.

\begin{figure}
	\centering
	\includegraphics[width=\columnwidth]{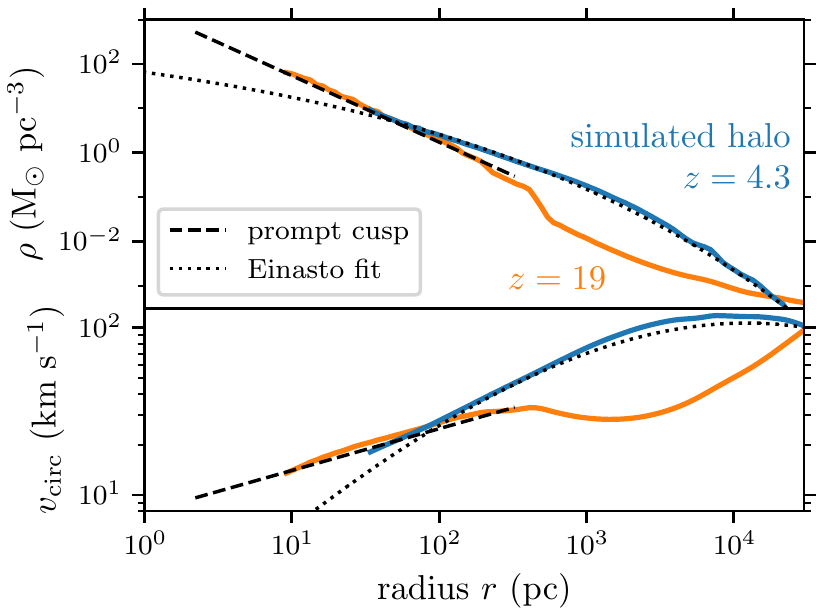}
	\caption{Radial profiles of the density $\rho$ (top) and the circular orbit velocity $\vc$ (bottom) for a \chg{simulated} halo arising in a 3.5~keV WDM scenario.
    Shortly after collapse (orange), the density profile closely matches the $\rho\propto r^{-1.5}$ prompt cusp that is predicted from the initial conditions (dashed line; not a fit). Later (blue), accretion has built up the density at radii above about 100~pc to the extent that it closely matches the Einasto profile, $\diff\log\rho/\diff\log r\propto -r^{\alpha}$, with $\alpha=0.17$ (dotted curve). However, the prompt cusp persists down to the simulation's resolution limit, significantly boosting $\vc$ at radii below 100~pc compared to what
    \chg{the Einasto profile would suggest.
    This cusp is only expected to flatten into a finite-density core below about 2~pc (see footnote~\ref{foot:core}).}
    }
	\label{fig:profile}
\end{figure}

In warm dark matter (WDM) models, the dark matter consists of keV-scale particles that were once in thermal contact with the Standard Model plasma. The thermal streaming motion then smooths the initial density field on greater-than-kiloparsec
scales \citep[e.g.][]{2001ApJ...556...93B}. Such smoothing can also arise from nonthermal production mechanisms, where the dark matter is initially relativistic \citep[e.g.][]{2019PrPNP.104....1B,2019PhRvD.100l3520M,2020JCAP...08..045B}, or in other contexts \citep[e.g.][]{2013IJMPA..2830042K,2019JCAP...10..055A,2021PhRvD.103d3526R}.
But the smoothing scale sets the mass scale of the peaks in the initial density field, which in turn determines the masses of the prompt density cusps.
For keV-scale WDM, prompt cusps can be of order $10^7$~M$_\odot$, massive enough to detectably influence the orbits of stars.\footnote{\chg{In contrast, prompt cusps in cold dark matter models may be Earth-mass \citep[][]{2022arXiv220911237D}.}}
In this article, we propose the prompt cusps of galactic haloes as a new probe of WDM and similar scenarios.

Figure~\ref{fig:profile}, adapted from \citet{2023MNRAS.518.3509D}, illustrates the scale of a WDM prompt cusp. We plot the density and circular velocity profiles for a halo that arose in a 3.5~keV WDM scenario. The halo develops a prompt $\rho\propto r^{-1.5}$ density cusp (dashed line) after its formation at redshift $z\sim 20$ (orange). By $z\sim 4$ (blue), the halo has built up a density profile of the \citet{1965TrAlm...5...87E} form (dotted curve),
which is the standard fitting form adopted in simulations of collisionless dark matter \citep[e.g.][]{2004MNRAS.349.1039N,2010MNRAS.402...21N,2020Natur.585...39W,2023MNRAS.519.3292D}. Nevertheless, the prompt cusp persists unaltered at the system's center to the extent that the simulation can resolve it.\footnote{\chg{Figure~\ref{fig:profile} shows} the simulated halo W1 of \citet{2023MNRAS.518.3509D}, scaled to match a 3.5~keV WDM cosmology as described in that article. \chg{The minimum resolved radius is taken to be five times the force-softening length and encloses about 900 particles at $z=19$.}} Relative to the dark matter halo structure that is normally assumed, this prompt cusp significantly boosts the halo's circular orbit velocity at radii below about 100~pc.

Previous probes of WDM and similar scenarios are based on the suppression of initial density variations at scales below the smoothing scale in the initial conditions. For example, the Lyman~$\alpha$ forest \citep[e.g.][]{2017PhRvD..96b3522I,2021MNRAS.502.2356G} can constrain WDM by showing that small-scale density variations are present.
Other probes, such as strong gravitational lensing \citep[e.g.][]{2020MNRAS.492.3047H,2020MNRAS.491.6077G}, the abundance of satellite galaxies \citep[e.g.][]{2021JCAP...08..062N,2021PhRvL.126i1101N}, and perturbations to stellar streams \citep[e.g.][]{2021MNRAS.502.2364B}, constrain WDM by finding the haloes that form from such small-scale density variations.
These probes produce lower limits on the WDM particle mass ranging from about 2~keV to about 6~keV, and recent combined analyses yield limits of $m_\chi>6$~keV \citep{2021MNRAS.506.5848E} and $m_\chi>10$~keV \citep{2021ApJ...917....7N}.

Compared to these ideas, prompt $\rho\propto r^{-1.5}$ cusps represent an entirely new probe. This article explores the impact of WDM prompt cusps on the dwarf galaxies of the Local Group. We find that WDM models with masses in the 2 to 6~keV range could alter the kinematics of these galaxies at detectable levels.


\section{Prompt cusps in WDM scenarios}\label{sec:cusps}

We begin with the \chg{linear} matter power spectrum $P_\mathrm{CDM}(k)$ for a cold dark matter (CDM) scenario computed using the \textsc{CLASS} code \citep{2011JCAP...07..034B} at
\chg{the (arbitrarily chosen) redshift $z=36$. By}
the prescription of \citet{2002MNRAS.333..544H} \citep[see also][]{2005PhRvD..71f3534V},
\chg{the WDM power spectrum is $P_\mathrm{WDM}(k)=T^2(k)P_\mathrm{CDM}(k)$, where
\begin{align}\label{transfer}
    T(k) = [1+(\lcut k)^{2\nu}]^{-5/\nu}, \ \ \ \nu=1.12.
\end{align}
}Here, $\lcut$ is the characteristic scale below which free streaming suppresses density variations in the dark matter and is given by
\begin{equation}\label{cut}
    \lcut = 0.070\,
    (m_\chi/\text{keV})^{-1.11}
    (\Omega_\chi h^2/0.12)^{0.11}
    \,\text{Mpc},
\end{equation}
where $m_\chi$ is the dark matter particle mass, $\Omega_\chi$ is the dark matter density today in units of the critical density, and
$h$ is the Hubble parameter.
\chg{We will also need the late-time linear growth function
\begin{equation}
    D(z) = (1+z)^{-1}\,_2F_1(1/3,1;11/6;-\rhoM/\rhoL)
\end{equation}
\citep[e.g.][]{2011JCAP...10..010B}, where
$_2F_1$ is the hypergeometric function and $\rhoM/\rhoL= (1+z)^{3}\OmegaM/\OmegaL$ is the ratio of matter to dark energy density as a function of $z$. We fix $\Omega_\chi h^2=0.12$, $\OmegaM=0.31$, and $\OmegaL=0.69$ \citep{2020A&A...641A...6P}.}


We evaluate $P_\mathrm{WDM}(k)$ for several values of $m_\chi$ and use the procedure of \citet{2022arXiv220911237D} to sample the prompt cusp distribution in each case. This procedure
begins by using 
Gaussian statistics
to randomly sample $10^6$ \chg{peaks in the linear density field} distributed in the amplitude $\delta\equiv\delta\rho/\bar\rho$, the curvature $\nabla^2\delta$, and the ellipticity $e$ and prolateness $p$ of the tidal tensor at the peak.\footnote{\chg{The $(\delta,\nabla^2\delta)$ distribution comes from \citet{1986ApJ...304...15B} and is subject to the peak constraint, while the $(e,p)$ distribution comes from \citet{2001MNRAS.323....1S} and is only conditioned on $\delta$. It can be verified numerically that the latter is not altered significantly if the peak constraint is imposed, however.}}
We next \chg{evaluate a peak's collapse redshift $\zc$ by inverting
\begin{equation}
    D(\zc) = D(z)\,\delta_\mathrm{ec}(e,p)/\delta,
\end{equation}
where $\delta_\mathrm{ec}(e,p)$ is the approximation given by \citet{2001MNRAS.323....1S} of the linear threshold for ellipsoidal collapse and $z=36$ is the redshift of $P_\mathrm{WDM}(k)$.}
\chg{Given $\zc$ and the characteristic size $q\equiv|\delta/\nabla^2\delta|^{1/2}$ of each peak,
the simulations of \citet{2023MNRAS.518.3509D} show that the resulting prompt cusp has a $\rho= A r^{-1.5}$ density profile extending out to radius $R$,\footnote{\label{foot:core} \chg{Above $R$, the density can (but does not necessarily) drop below $A r^{-1.5}$.} There is also an interior radius at which prompt cusps are expected to give way to central cores \citep{2023MNRAS.518.3509D}. Using the initial thermal velocity distribution given by \citet{2001ApJ...556...93B}, we verified that core radii in WDM scenarios are about \chg{$R/200$, negligibly small} in the current context.} where
\begin{align}
    A &\simeq 24\rhoMo(1+\zc)^{1.5}q^{1.5}, \\ R &\simeq 0.11 (1+\zc)^{-1} q.
\end{align}
Here, $\rhoMo\simeq 39$~M$_\odot$kpc$^{-3}$ is the comoving matter density.}
The coefficient $A$ predicted in this way matches \citet{2023MNRAS.518.3509D} simulation results to within about 10 per cent.
\chg{This density profile should be viewed as a lower limit, as mass accreted later can overlap with and boost the density of the central cusp (see Fig.~\ref{fig:profile}).}

\begin{figure}
	\centering
	\includegraphics[width=\columnwidth]{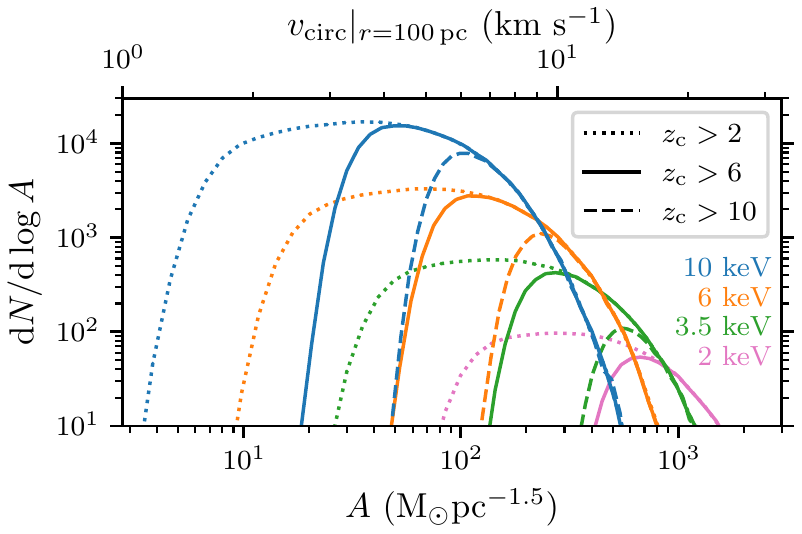}
	\caption{Average number of peaks in the initial density field within a $3\times 10^{12}$~M$_\odot$ mass of matter, logarithmically distributed \chg{in}
    the density coefficients $A$ of the $\rho=A r^{-1.5}$ prompt cusps that their collapse would produce. We consider four different WDM particle masses (colours), and we restrict to peaks that collapse by $z=10$ (dashed curves), $z=6$ (solid), and $z=2$ (dotted). Lower WDM particle masses lead to more pronounced prompt cusps (larger $A$), and earlier collapse also produces denser cusps.
    The upper axis shows the cusps' circular orbit velocities $\vc\propto r^{1/4}$ at the radius $r=100$~pc.}
	\label{fig:distA}
\end{figure}

For a range of WDM scenarios, Fig.~\ref{fig:distA} shows the average differential number $\diff N/\diff\log A$ of density peaks within a $3\times 10^{12}$~M$_\odot$ mass of matter, which is approximately the mass of the Local Group \citep{2022ApJ...928L...5B}.
These peaks are distributed in the density coefficients $A$ of the prompt cusps that result from their collapse. Generally, a smaller WDM particle mass $m_\chi$ leads to a larger smoothing scale for the initial density field (see Eq.~\ref{cut}) and hence more pronounced prompt cusps.
We will focus on galaxies that formed at high redshift, so the peak distributions in Fig.~\ref{fig:distA} are truncated according to their collapse times: the dotted curves indicate collapse redshift $\zc>2$, the solid curves indicate $\zc>6$, and the dashed curves indicate $\zc>10$. Earlier cusp formation is clearly associated with more pronounced prompt cusps (larger $A$ or $\vc$).





\section{Local Group dwarf galaxies}\label{sec:LG}

\begin{figure}
	\centering
	\includegraphics[width=\columnwidth]{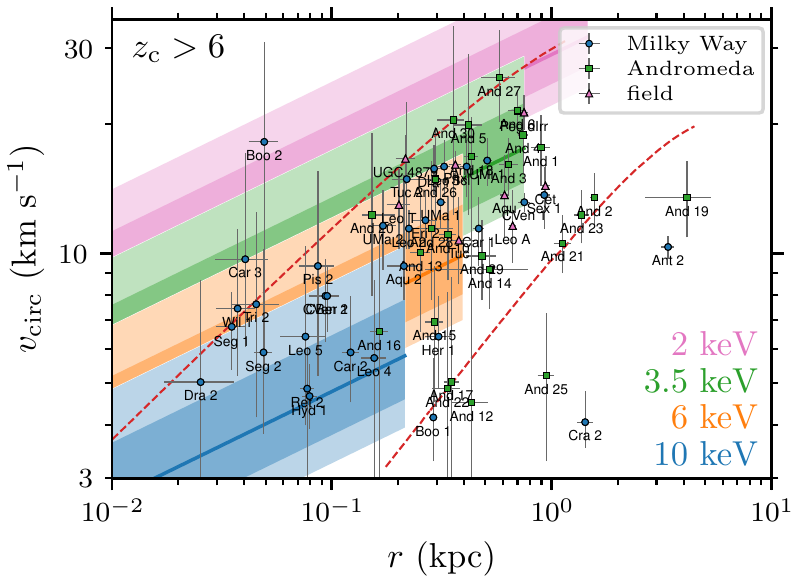}
	\caption{Comparison between Local Group dwarf galaxy kinematics and the prompt cusp kinematics predicted by WDM models. As points, we plot the observationally inferred circular orbit velocity $\vc$ at the half-light radius $r=\rh$ for Milky Way satellites (circles), Andromeda satellites (squares), and dwarf galaxies in the Local Group field (triangles). The bands show the median and the 68 and 95 per cent ranges of $\vc$ for the prompt cusps arising in four representative WDM models (colours). Since the smallest galaxies are likely to have formed before reionization, we consider only cusps that form by $z=6$. For each WDM model, we plot $\vc$ up to the median cusp radius $R$. The distributions that are partially occluded can be extrapolated linearly, as $\vc\propto r^{1/4}$ for a prompt cusp. Generally, with respect to a WDM model's prompt cusp distribution, galaxies with excessively low $\vc$ present a challenge to the model while galaxies with excessively high $\vc$ do not. For comparison, the dashed lines enclose the region where dwarf galaxies are expected to lie in CDM scenarios, per \citet{2022MNRAS.511.6001E}.}
	\label{fig:vcirc-demo}
\end{figure}

We now compare the prompt cusps that arise in WDM scenarios with the observed kinematics of the dwarf galaxies of the Local Group. For each galaxy, Figure~\ref{fig:vcirc-demo} shows the circular orbit velocity $\vc$ at the half-light radius $\rh$. We follow
\citet{2010MNRAS.406.1220W}
in estimating $\rh=4/3 \Rh$ and $\vc=\sqrt{3}\sigma_\mathrm{los}$, where $\Rh$ is the projected half-light radius and $\sigma_\mathrm{los}$ is the line-of-sight velocity dispersion. These measured quantities are sourced from \citet{2012AJ....144....4M} (January 2021 version) with further data drawn from 
\citet{2016MNRAS.458L..59M,2016ApJ...818...40M,
2018ApJ...863...25M,2019MNRAS.487.2961M,2020MNRAS.491.3496C,2020A&A...635A.152T,2021MNRAS.505.5686C,2021ApJ...921...32J}.
The same figure also shows the ranges of circular orbit velocities
\begin{equation}
    \vc=\sqrt{8\pi G A/3}\, r^{1/4}
\end{equation}
associated with the prompt $\rho=A r^{-1.5}$ cusps arising for the WDM models considered in Section~\ref{sec:cusps} (see Fig.~\ref{fig:distA}). We restrict to $\zc>6$
because it is likely that the smallest galaxies formed prior to the epoch of reionization \citep[e.g.][]{2020MNRAS.498.4887B}.\footnote{Within the mass of the Local Group, the 2~keV, 3.5~keV, 6~keV, and 10~keV models yield on average 48, 440, 3300, and 21000 peaks that collapse by $\zc>6$, respectively (see Fig.~\ref{fig:distA}). There are 65 galaxies shown in Fig.~\ref{fig:vcirc-demo}. Thus, in the 2~keV case, not all galaxies shown could have formed by $z=6$, but the lower-mass galaxies most relevant to this study could have. For higher WDM masses, the peak count does not limit the fraction of known galaxies that could have formed by $z=6$.} Since accreted material can build density on top of a halo's prompt cusp (see Fig.~\ref{fig:profile}), dwarf galaxies that lie above the cusp distribution are not concerning. However, galaxies that lie below a WDM model's prompt cusp distribution present challenges to that model. For comparison, we also show (dashed lines) where \chg{\citet{2022MNRAS.511.6001E} predict that dwarf galaxies should lie in CDM scenarios, based on the \citet{1997ApJ...490..493N} density profile and simulations of its tidal evolution.}

\begin{figure}
	\centering
	\includegraphics[width=\columnwidth]{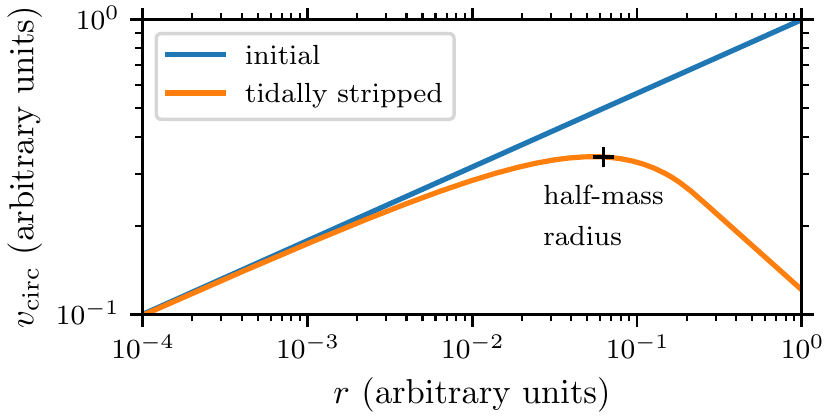}
	\caption{Circular velocity profile of a prompt cusp \chg{tidally stripped in the asymptotic limit} (orange). At the half-mass radius, $\vc$ is scaled by a factor of about 0.68 compared to the initial prompt cusp (blue).}
	\label{fig:vcirc-tidal}
\end{figure}

Figure~\ref{fig:vcirc-demo} suggests that prompt cusps may threaten WDM masses as high as 6~keV. However, there is a further consideration. The satellite dwarf galaxies have been modified by the tidal forces of their hosts, which can suppress their $\vc$ at larger radii. Figure~\ref{fig:vcirc-tidal} shows the circular velocity profile of a tidally stripped
\chg{$\rho\propto r^{-1.5}$ cusp, evaluated in the asymptotic limit}
using the model of \citet{2022arXiv220700604S}. Tidal stripping scales $\vc$ at the half-mass radius by a factor of 0.68.
\chg{Tidal forces truncate the stellar and dark matter distributions at the same radius, so
in the limit that a galaxy is severely stripped,
the half-light radius should lie close to the half-mass radius \citep[e.g.][]{2022MNRAS.511.6001E}}. Thus, we will assume that in this limit, $\vc$ at the half-light radius $\rh$ is scaled by the same 0.68 factor.
Outside of this limit, galaxies are generally smaller than their haloes, so $\vc$ \chg{at $\rh$} would be suppressed to a more modest degree. \chg{The outcome is a lower limit on $\vc$ for this reason, because the tidally truncated profile was evaluated in the asymptotic limit, and because we began with a prompt cusp profile not built up by any further accretion.}

\begin{figure}
	\centering
	\includegraphics[width=\columnwidth]{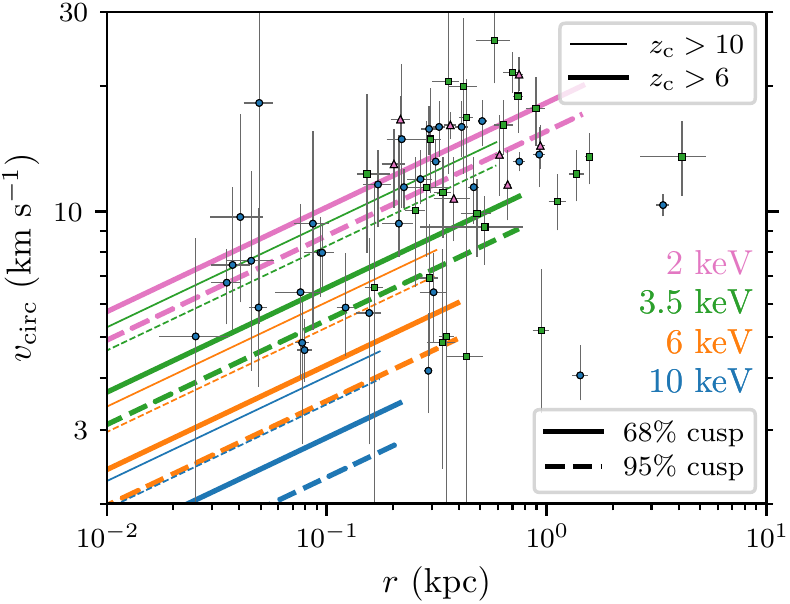}
	\caption{Compatibility of Local Group dwarf galaxy kinematics with the prompt cusps predicted by WDM models. As in Fig.~\ref{fig:vcirc-demo}, we plot the observationally inferred circular orbit velocity $\vc$ at the half-light radius $r=\rh$ for Local Group dwarf galaxies; symbols have the same meaning in both figures. The lines indicate the $\vc$ above which 95 per cent (solid) or 68 per cent (dashed) of the prompt cusps lie; different colors represent different WDM models, and we assume the cusps are maximally tidally stripped, scaling their $\vc$ by 0.68. \chg{For the thick lines,}
    we restrict the prompt cusp distribution to those that form before $z=6$, based on the likelihood that a low-mass galaxy formed before reionization. This assumption seriously threatens 3.5~keV WDM and places 6~keV WDM in tension with several satellite galaxies with half-light radii $\rh\gtrsim 0.3$~kpc.
    \chg{For the thin lines,}
    we truncate the distributions such that $\zc>10$ instead. This assumption places 6~keV WDM in much more serious tension with a range of galaxies.
    \chg{There is no thin line for the 2~keV WDM scenario}
    because only about four peaks would be expected to collapse within the Local Group by $z=10$.}
	\label{fig:vcirc}
\end{figure}

Figure~\ref{fig:vcirc} compares the Local Group circular velocities to the WDM predictions for maximally tidally stripped prompt cusps. For each WDM model, we now mark the $\vc$ above which 68 per cent (solid lines) and 95 per cent (dashed lines) of the prompt cusps lie. \chg{The thick lines denote the prompt cusp distributions truncated such that $\zc>6$, as above.}
The 2~keV WDM scenario is clearly ruled out, and the 3.5~keV scenario is in serious tension with a number of satellite galaxies with $\rh\gtrsim 100$~pc. Even the 6~keV model is threatened by Bo{\"o}tes I and several Andromeda satellites. We remark, however, that these latter galaxies are of intermediate size, so the assumption that they formed before reionization may be less well motivated. Additionally, larger dwarf galaxies tend to be more susceptible to modification of their central cusps by baryonic feedback \citep[e.g.][]{2019MNRAS.484.1401R,2020MNRAS.491.4523F,2023MNRAS.518.5356L}.

\chg{The thin lines in Fig.~\ref{fig:vcirc}} show prospects for constraining WDM if the relevant prompt cusp formation times could be determined to be earlier. We now truncate the distribution at $\zc>10$. For example, perhaps some of the observed galaxies must have formed before a significant fraction of the gas reionized. Under this assumption, the 6~keV WDM model is in much more serious tension with a range of dwarf galaxies, although the 10~keV model remains safe.

\begin{figure*}
	\centering
	\includegraphics[width=\linewidth]{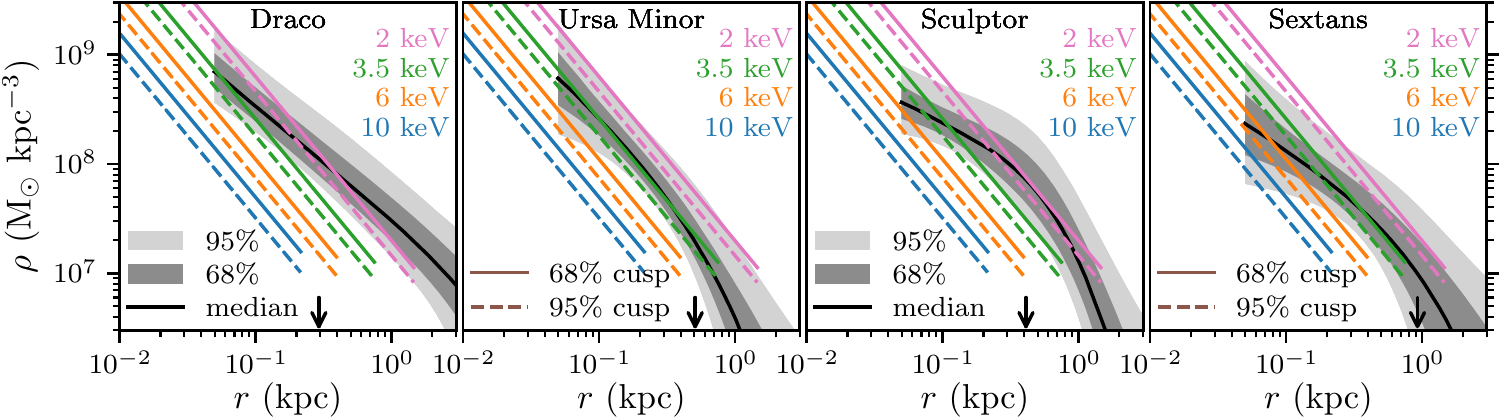}
	\caption{Compatibility of the resolved kinematics of several classical dwarfs (different panels) with the prompt cusps predicted by WDM models. We plot the density profiles inferred by \citet{2020ApJ...904...45H} in black (for the median) and different levels of gray (for the 68 per cent and 95 per cent confidence intervals). For each of four WDM models (colours), we plot the $\rho=A r^{-1.5}$ prompt cusps that are less dense than 68 per cent (solid lines) and 95 per cent (dashed lines) of the cusp distribution, where this distribution is truncated such that $\zc>6$. 2~keV WDM is likely incompatible with these galaxies' internal kinematics, while 3.5~keV WDM is in mild tension. To connect this picture with Fig.~\ref{fig:vcirc}, we mark the half-light radius $\rh$ with an arrow.}
	\label{fig:testprof}
\end{figure*}

For some galaxies, we can also search for signs of a prompt cusp within their detailed internal kinematics. Figure~\ref{fig:testprof} shows the density profiles of four classical dwarfs, as inferred by \citet{2020ApJ...904...45H}. For comparison, we also show the $\rho=A r^{-1.5}$ prompt cusp above which 68 per cent (solid lines) and 95 per cent (dashed lines) of the prompt cusps lie, with the distribution truncated such that $\zc>6$ again. Since we are focused on the deep interiors of these haloes, we do not include any tidal stripping. Based on Fig.~\ref{fig:testprof}, 2~keV WDM is likely incompatible with the kinematics of the Sextans and Sculptor dwarfs, while 3.5~keV WDM is only in mild tension.
As we discussed above, one complicating factor is that larger dwarfs such as these are more susceptible to cusp modification through baryonic feedback. However, this effect is likely small here: the galaxies we selected have had minimal star formation activity in the last 10 Gyr \citep{2009ApJ...703..692L,2014ApJ...789..147W}.

We also mark the half-light radius $\rh$ of each of the dwarf galaxies in Fig.~\ref{fig:testprof} with an arrow. Evidently, the deep interiors of these objects can be far more constraining than the kinematics at $\rh$ alone. This outcome suggests that limits on WDM could be significantly improved by considering the internal kinematics of more galaxies, particularly those that already drive the most serious tensions in Fig.~\ref{fig:vcirc}.

\section{Outlook}

The prompt $\rho\propto r^{-1.5}$ density cusps expected to lie at the centre of every halo and subhalo present a new test of WDM. They arise from smooth peaks in the initial density field and are more pronounced in models where the dark matter's streaming motion smooths the density field on larger scales. We showed that WDM models with masses in the 2 to 6~keV range (and lower) may be in tension with the observed kinematics of Local Group dwarf galaxies because no sign is present of the prompt cusps that should arise in these models.

We employed two main approaches. First, we checked whether the prompt cusps of WDM models are compatible with the overall velocity dispersion of each known dwarf. Second, we explored the compatibility of WDM prompt cusps with the deep internal kinematics of several classical dwarfs for which these kinematics have been resolved. While the former approach was more constraining on WDM models, the value of the latter approach could improve with access to the internal kinematics of lower-mass galaxies.

The density of a prompt cusp is highly sensitive to its formation time, which is \chg{when} it condensed out of the smooth initial mass distribution. Earlier-forming prompt cusps are denser and hence produce stronger observational signatures. Another major avenue toward refining prompt cusp-based limits on WDM models is to clarify the ages of the prompt cusps that lie at the centres of Local Group dwarfs.
We employed assumptions connected to the notion that the smallest galaxies are likely to have formed before the epoch of reionization, but other approaches are also possible.
One approach could be to explore, in models \citep[e.g.][]{2019PhRvD.100b3523D} or numerical simulations, which initial density peaks tend to be associated with prompt cusps whose haloes match the known properties \chg{and environment of Local Group dwarfs.
For example, Local Group haloes, and especially satellite haloes, originate from a biased initial volume, an effect that would raise their formation redshifts.}
Another possibility is to employ constraints from galactic archaeology on the ages of these galaxies.

Also, in our comparison with the Local Group dwarf galaxy population, we conservatively assumed that tidal stripping has modified the structures of their prompt cusps to a maximal degree. This assumption is likely not valid for many of the galaxies, though. Constraints on WDM can be further tightened by accounting for a level of tidal stripping that is consistent with the observationally inferred orbits of Local Group satellite galaxies.

Beyond galactic kinematics, prompt cusps should also produce signatures in other probes of dark matter structure. In WDM models, prompt cusps could affect strong gravitational lensing and perturbations to stellar streams. These probes have already been employed to constrain WDM through its suppression of the abundance of low-mass haloes. Particularly in the context of strong lensing, WDM's impact on the internal structures of haloes has also been accounted for, but this accounting was based on simulations that did not resolve the central prompt cusps \citep{2016MNRAS.455..318B}. Accounting for these, as we do here, could strengthen the constraints on WDM models.

\section*{Acknowledgements}

The author thanks Simon White for helpful discussions and comments on the manuscript, Rapha{\"e}l Errani for supplying a copy of the Local Group dwarf galaxy data used in Figs. \ref{fig:vcirc-demo} and~\ref{fig:vcirc}, and Eiichiro Komatsu and Fabian Schmidt for comments on the manuscript.

\section*{Data Availability}
 
No new data were generated or analysed in support of this research.



\bibliographystyle{mnras}
\bibliography{main}



\appendix



\label{lastpage}
\end{document}